\documentclass[preprint,aps]{revtex4}

\usepackage{graphicx}
\usepackage{dcolumn}
\usepackage{bm}
\usepackage{epsf}

\newcommand{\beq}{\begin{equation}}
\newcommand{\eeq}{\end{equation}}
\newcommand\beqa{\begin{eqnarray}}
\newcommand\eeqa{\end{eqnarray}}
\newcommand\bea{\begin{array}}
\newcommand\eea{\end{array}}
\newcommand{\nn}{\nonumber}
\newcommand{\neqa}{\nonumber\end{eqnarray}}
\newcommand{\la}{\label}
\newcommand{\noi}{\noindent}
\newcommand{\eq}[1]{eq.(\ref{#1})}

\newcommand{\ur}[1]{(\ref{#1})}

\newcommand{\Tr}{{\rm Tr}}
\newcommand{\Det}{{\rm Det}}
\newcommand{\half}{\frac{1}{2}}
\renewcommand{\d}{\partial}

\newcommand{\D}{r_{12}}
\newcommand{\bv}{\overline{{\rm v}}}
\renewcommand{\v}{{\rm v}}

\begin{document}

\title{Quantum weights of monopoles and calorons with non-trivial holonomy}
\author{\bf Dmitri Diakonov}
\vskip 0.5true cm

\affiliation{Thomas Jefferson National Accelerator Facility, Newport News, VA 23606, USA\\
NORDITA, Blegdamsvej 17, DK-2100 Copenhagen, Denmark\\
St. Petersburg Nuclear Physics Institute, Gatchina, 188 300, St. Petersburg, Russia}

\date{July 29, 2004}

\vskip 1true cm

\begin{abstract}
Functional determinant is computed exactly for quantum
oscillations about periodic instantons with non-trivial values of
the Polyakov line at spatial infinity (or holonomy). Such instantons
can be viewed as composed of the Bogomolnyi--Prasad--Sommerfeld (BPS) 
monopoles or dyons. We find the weight or the probability with which 
dyons occur in the pure Yang--Mills partition function. It turns out
that dyons experience quantum interactions having the familiar
``linear plus Coulomb" form but with the ``string tension''
depending on the holonomy. We present an argument that at temperatures below 
the critical one computed from $\Lambda_{\rm QCD}$, trivial holonomy 
becomes unstable, with instantons ``ionizing'' into separate dyons. 
It may serve as a microscopic mechanism of the confinement-deconfinement 
phase transition~\cite{F1}.
\end{abstract}



\maketitle

\section{Introduction}
Several years ago a new self-dual solution of the Yang--Mills equations at
non-zero temperatures has been found independently by Kraan and van Baal~[2] 
and Lee and Lu~[3]. I shall call them for short the KvBLL calorons. In the case
of the simplest $SU(2)$ gauge group to which I restrict myself in this paper,
the KvBLL caloron is characterized by 8 parameters or collective coordinates,
as it should be according to the general classification of the self-dual
solutions with a unity topological charge. The most interesting feature of
the KvBLL calorons is that they can be viewed as composed of two ``constituent''
dyons; one is the standard BPS monopole~[4,5] and the other is the so-called 
Kaluza--Klein monopole~[6]. I denote them as $M,L$ dyons; explicitly
their fields can be found {\it e.g.} in the Appendix of Ref.~[7]. 

The collective coordinates or the moduli space of the 
KvBLL caloron can be chosen in various ways, however the physically most 
appealing choice is the six coordinates of the two dyons' centers in space 
$\overrightarrow z_{1,2}$, and two compact ``time'' variables. When the spatial 
separation of two constituent dyons $\D=|\overrightarrow z_1-\overrightarrow z_2|$ 
is larger than the compactification circumference $1/T$ the caloron action 
density becomes static and is reduced to the sum of the static action densities 
of the two dyons. At $\D T\leq 1$ the two dyons merge, and the action 
density becomes a time-dependent $4d$ lump, see Fig.~1.

\begin{figure}[ht]
\centerline{\epsfxsize=12cm\epsfbox{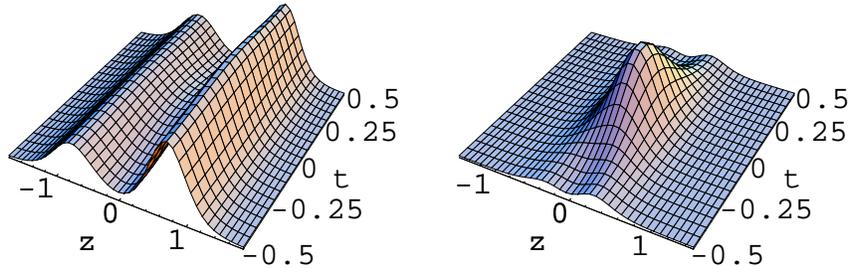}}   
\caption{The action density of the KvBLL caloron as function
of $z,t$ at fixed $x=y=0$, with the asymptotic value of $A_4$ at spatial
infinity $\v=0.9\pi T,\; \bv=1.1\pi T$. It is periodic in the $t$ direction.
At large dyon separation the density becomes static (left, $\D=1.5/T$).
As the separation decreases the action density becomes more like a $4d$
lump (right, $\D=0.6/T$). In both plots the dyons are centered at
$z_1= -\v\,\D/2\pi T,\;z_2=\bv\,\D/2\pi T,\;x_{1,2}=y_{1,2}=0$.
The axes are in units of temperature T.\label{fig:adp1}}
\end{figure}

We use the gauge freedom to choose the gauge where the $A_4$ component of the
Yang--Mills field is static and diagonal at spatial infinity, $A_4\to \half\tau^3\v,
\;\v\in [0,2\pi T]$. The Polyakov line or the holonomy at spatial infinity is then
\beq
\half\,\Tr\,L=\half\,\Tr\,{\rm P}\,\exp\left(i\int_0^{1/T}dt\,A_4\right)
\to \cos\frac{\v}{2T}\,\in [-1,1].
\la{TrL}\eeq 
At the end points ($\v=0,\,2\pi T$) the holonomy belongs to the group center,
$\half\Tr L=\pm 1$, and is said to be `trivial'. In this case the KvBLL caloron
is reduced to the standard periodic instanton, also called the Harrington--Shepard 
caloron~[8].

It has been known since the work of Gross, Pisarski and Yaffe~[9] that gauge
configurations with non-trivial holonomy, $\half\Tr L\neq\pm 1$, are strongly
suppressed in the Yang--Mills partition function. Indeed, the 1-loop effective 
action obtained from integrating out fast varying fields where one keeps all 
powers of $A_4$ but expands in (covariant) derivatives of $A_4$ has the form~[10]
\beqa
\nn
S_{\rm eff}&= &\int\!d^4x\,\left[P(A_4)+{\bf E}^2f_E(A_4)+{\bf B}^2f_B(A_4)
+ {\rm higher\; derivatives}\right],\\
\la{Tpot}
P(A_4)&=&\left.\frac{1}{3T(2\pi)^2}\v^2(2\pi T-\v)^2\right|_{{\rm  mod}\; 2\pi T},
\qquad \v=\sqrt{A_4^aA_4^a},
\eeqa
where the perturbative potential energy term $P(A_4)$ has been known for a long
time~[9,11], see Fig.~2. The zeros of the potential energy correspond to 
$\half\Tr\,L= \pm 1$, {\it i.e.} to the trivial holonomy. If a dyon has $\v\neq 2\pi T n$ 
at spatial infinity the potential energy is positive-definite and proportional to the $3d$
volume. Therefore, dyons and KvBLL calorons with non-trivial holonomy seem
to be strictly forbidden: quantum fluctuations about them have an unacceptably
large action. 

\begin{figure}[ht]
\centerline{\epsfxsize=6cm\epsfbox{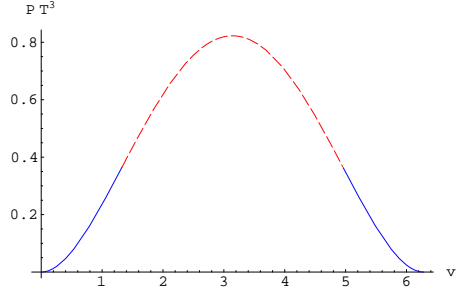}}   
\caption{Potential energy as function of $\v/T$.
Two minima correspond to $\half\Tr L=\pm 1$, the maximum corresponds to $\Tr L=0$.
The range of the holonomy where dyons experience repulsion is shown in dashing.\label{fig:pot_en}}
\end{figure}

Meanwhile, precisely these objects determine the physics of the
supersymmetric YM theory where in addition to gluons there are gluinos,
{\it i.e.} Majorana (or Weyl) fermions in the adjoint representation. Because
of supersymmetry, the boson and fermion determinants about dyons cancel exactly,
so that the perturbative potential energy \ur{Tpot} is identically zero for
all temperatures, actually in all loops. Therefore, in the supersymmetric theory
dyons are openly allowed. [To be more precise, the cancellation occurs when
periodic conditions for gluinos are imposed, so it is the compactification
in one (time) direction that is implied, rather than physical temperature
which requires antiperiodic fermions.] Moreover, it turns out~[12] 
that dyons generate a non-perturbative potential having a
minimum at $\v=\pi T$, {\em i.e.} where the perturbative potential would
have the maximum. This value of $A_4$ corresponds to the holonomy
$\Tr\,L=0$ at spatial infinity, which is the ``most non-trivial'';
as a matter of fact $<\!\Tr\,L\!>=0$ is one of the confinement's requirements.

In the supersymmetric YM theory there is a non-zero gluino condensate whose correct 
value is reproduced by saturating it by the $L,M$ dyons' zero fermion modes~[12]. 
On the contrary, the saturation of the (square of) gluino condensate by instanton 
zero modes gives the wrong result, namely $\sqrt{4/5}$ that of the correct value~[13]. 
Recently it has been observed~[7] that the square of the gluino condensate 
must be computed not in the instanton background but in the background of exact 
solutions ``made of'' $LL,\,MM$ and $LM$ dyons. The first two are the double-monopole 
solutions and the last one is the KvBLL caloron. As the temperature goes to zero, 
the $LL$ and $MM$ solutions have locally vanishing fields, whereas the KvBLL $LM$ 
solution reduces to the instanton field up to a locally vanishing difference. 
Therefore, naively one would conclude that the dyon calculation of the gluino 
condensate, which is a local quantity, should be equivalent to the instanton one, 
but it is not. The fields vanishing as the inverse size of the system have
a finite effect on such a local quantity as the gluino condensate! This is quite unusual. 
The crucial difference between the (wrong) instanton and the (correct) dyon calculations 
is in the value of the Polyakov loop, which remains finite. In the $N\!=\!1$ SUSY
theory, as one increases the compactification circumference $1/T$, the average
$<\!A_4\!>=\pi T$ at infinity decreases, however the theory always remains in the Higgs
phase, in a sense. Instantons do not satisfy this boundary conditions whereas
dyons and calorons with the non-trivial holonomy do satisfy them. 

In the supersymmetric YM theory configurations having $\Tr\,L=0$ 
at infinity are not only allowed but dynamically preferred as compared to those 
with $L=\pm 1$. In a non-supersymmetric theory it looks as if it is the opposite. 

Nevertheless, it has been argued in Ref.~[14] that the perturbative
potential energy \ur{Tpot} which forbids individual dyons in the pure YM
theory might be overruled by non-perturbative contributions of an {\em ensemble}
of dyons. For fixed dyon density, their number is proportional to the
$3d$ volume and hence the non-perturbative dyon-induced potential as function
of the holonomy (or of $A_4$ at spatial infinity) is also proportional to the volume.
It may be that at temperatures below some critical one the non-perturbative potential
wins over the perturbative one so that the system prefers $<\!\Tr\,L\!>=0$.
This scenario could then serve as a microscopic mechanism of the confinement-deconfinement
phase transition~[14]. It should be noted that the KvBLL calorons and
dyons seem to be observed in lattice simulations below the phase transition
temperature~[15,16,17]. 

To study this possible scenario quantitatively, one first needs to find out
the quantum weight of dyons or the probability with which they appear
in the Yang--Mills partition function. At the second stage, one has to consider
the statistical mechanics of the $L,M,\overline L,\overline M$ dyons for fixed
$A_4$ at infinity and find the free energy of the system as function of 
$\v\equiv \left.\sqrt{A_4^aA_4^a}\right|_{|\vec x|\to\infty}$. Finally, one
has to study this free energy as function of $\v$ at different temperatures,
to see what value of $\v$ (equivalent to the holonomy according to the formula
$\half\Tr\,L=\cos(\v/2T)$) is preferred by the theory.  

Unfortunately, the single-dyon measure 
is not well defined: it is too badly divergent in the infrared region owing to 
the weak (Coulomb-like) decrease of the fields. What makes sense and is finite, 
is the quantum determinant for small oscillations about the KvBLL caloron made of
two dyons with zero combined electric and magnetic charges. Knowing the weight
of the electric- and magnetic-neutral KvBLL caloron one can read off the
individual $L,M$ dyons' weights and their interaction as one moves the two dyons
apart.

The problem of computing the effect of quantum fluctuations about a caloron with
non-trivial holonomy is of the same kind as that for ordinary instantons (solved
by 't Hooft~[18]) and for the standard Harrington--Shepard caloron (solved
by Gross, Pisarski and Yaffe~[9]) being, however, technically much more difficult.
I remind the results of the above two calculations in the next two sections.
In Section 4 I report on the very recent result for the KvBLL caloron~[1].
Remarkably, the quantum weight of the KvBLL caloron can be computed {\em exactly}.
It becomes possible because we are able to construct the exact propagator of spin-0, 
isospin-1 field in the KvBLL background, which is some achievement by itself. 

It turns out from the exact calculation of the KvBLL weight that dyons experience
a familiar ``linear plus Coulomb'' interaction at large separations. That is why the 
individual dyon weight is ill-defined: their interaction grows with the separation.
The sign of the interaction depends critically 
on the value of the holonomy. If the holonomy is not too far from the trivial such that 
$0.787597\!<\!\half|\Tr L|\!<\!1$, corresponding to the positive second derivative 
$P''(\v)$ (see Fig.~2) the $L$ and $M$ dyons experience a linear {\em attractive} potential. 
Integration over the separation $\D$ of dyons inside a caloron converges. 
We perform this integration in Section 6 assuming calorons are in the ``atomic''
phase, estimate the free energy of the neutral caloron gas 
and conclude that the trivial holonomy ($\v=0,2\pi T$) is unstable at temperatures below
$T_c=1.125\,\Lambda_{\overline{\rm MS}}$, despite the perturbative potential energy $P(\v)$. 
In the complementary range $\half|\Tr L|\!<\!0.787597$, the second derivative $P''(\v)$ 
is negative, and dyons experience a strong linear-rising {\em repulsion}. It means that for 
these values of $\v$, integration over the dyon separations diverges: calorons with holonomy
far from trivial ``ionize'' into separate dyons. 

\section{Ordinary instantons at $T=0$}
The usual Belavin--Polyakov--Schwartz--Tyupkin instanton~[19] has the field 
\beq\nn
A_\mu=A_\mu^at^a
=\frac{-i\rho^2U[\sigma^-_\mu(x-z)^+-(x-z)_\mu]U^\dagger}
{(x-z)^2[\rho^2+(x-z)^2]},\qquad \sigma_\mu^\pm=({\bf 1},\pm i{\overrightarrow \tau}).
\eeq
The moduli or parameter space is described by the center $z_\mu$ (4), size $\rho$ (1), 
and orientation $U$ (3). The action density $\Tr\,F_{\mu\nu}^2$ is $O(4)$ symmetric,
see Fig.~3.

As it is well known~[18], the calculation of the 1-loop quantum weight
of a Euclidean pseudoparticle consists of three steps: i) calculation of the
metric of the moduli space or, in other words, calculation of the Jacobian composed
of zero modes, needed to write down the pseudoparticle measure in terms of its
collective coordinates, ii) calculation of the functional determinant for
non-zero modes of small fluctuations about a pseudoparticle, iii) calculation of
the ghost determinant resulting from background gauge fixing in the previous step.
In fact, for self-dual fields problem ii) is reduced to iii) since for such fields 
$\Det(W_{\mu\nu})= \Det(-D^2)^4$, where  $W_{\mu\nu}$ is the quadratic form for spin-1, 
isospin-1 quantum fluctuations and $D^2$ is the covariant Laplace operator for 
spin-0, isospin-1 ghost fields~[20]. Symbolically, one can write
\beq
{\rm pseudoparticle\;weight}=\!\!\int\!d({\rm collective\; coordinates})
\cdot{\rm  Jacobian}\cdot \Det^{-1}(-D^2),
\la{symbolic}\eeq
The functional determinant is normalized to the free one (with zero background fields) 
and UV regularized by the standard Pauli--Villars method. 

The 1-loop quantum weight of the BPST instanton has been computed by 't Hooft~[18].
If $\mu$ is the Pauli--Villars mass, {\it i.e.} the UV cutoff, and $g^2(\mu)$ is
the gauge coupling given at this cutoff, the instanton weight is
\beqa\nn
&& e^{-\frac{8\pi^2}{g^2(\mu)}}\int \!d^4z\,d^3U\,\frac{d\rho}{\rho^5}\,
(\mu\rho)^8\frac{1}{4\pi^2}\left(\frac{8\pi^2}{g^2(\mu)}\right)^4\,
\left\{C_0\,(\mu\rho)^{-\frac{2}{3}}\;\left[=\!\Det^{-1}(-D^2)\right]\right\} \\
\nn
&& C_0=\exp\left[-\frac{2\gamma_E}{3}+\frac{16}{9}-\frac{\log 2}{3}-\frac{2\log (2\pi)}{3}
+\frac{4\zeta'(2)}{{\pi }^2}\right]=0.64191.
\eeqa
The last factor (in the curly brackets) is due to the regularized small-oscillation
determinant; all the rest is actually arising from the 8 zero modes. The combination
$\mu^{\frac{22}{3}}e^{-\frac{8\pi^2}{g^2(\mu)}}=\Lambda^{\frac{22}{3}}$ is the scale 
parameter which is renormalization-invariant at one loop. Since the action density is 
$O(4)$ symmetric, the quantum weight depends only on the dimensionless quantity
$\rho\Lambda$ where $\rho$ is the instanton size, and even this dependence follows 
trivially from the known renormalization properties of the theory. Therefore, only
the overall numerical constant $C_0$ is the non-trivial result of the calculation.
The prefactor $g(\mu)^{-8}$ is not renormalized at one loop.

At two loops the instanton weight can be obtained without further calculations~[21] 
if one demands that it should be invariant under the simultaneous change of the cutoff 
$\mu$ and $g^2(\mu)$ given at this cutoff, such that the 2-loop scale parameter
\beqa\nn
\Lambda&=&\mu\,\exp\left(-\frac{8\pi^2}{b_1g^2(\mu)}\right)\,
\left(\frac{16\pi^2}{b_1g^2(\mu)}\right)^{\frac{b_2}{2b_1^2}}\,
\left[1+O\left(g^2(\mu)\right)\right],\\
\nn
&&b_1= \frac{11}{3}N,\quad b_2= \frac{34}{3}N^2,
\eeqa
remains fixed. The 2-loop instanton weight computed from this requirement is~[22]
\beqa\nn
&& \int \!d^4z\,dU\,\frac{d\rho}{\rho^5}\,\frac{C_0}{4\pi^2}\,
\beta(\rho)^{4}\,\exp\left[-\beta^{\rm II}(\rho)
\!+\!\left(4\!-\!\frac{b_2}{2b_1}\right)\!\frac{b_2}{2b_1}
\frac{\ln\beta(\rho)}{\beta(\rho)}+O\left(\frac{1}{\beta(\rho)}\right)\right],\\
\nn
&& \beta(\rho)= b_1\ln\frac{1}{\rho\Lambda},\qquad
\beta^{\rm II}(\rho)= \beta(\rho)+\frac{b_2}{2b_1}\ln\frac{2\beta(\rho)}{b_1}.
\eeqa

\section{Quantum weight of the periodic instanton with trivial holonomy}  
The Harrington--Shepard caloron~[8] is an immediate generalization of the ordinary 
Belavin--Polyakov--Schwartz--Tyupkin instanton~[19], continued by periodicity in the
time direction. The 1-loop quantum weight of the periodic instanton with the trivial
holonomy has been computed by Gross, Pisarski and Yaffe~[9]. The weight is a function of the
instanton size $\rho$, temperature $T$ and, after the renormalization, of the scale
parameter $\Lambda$.

\begin{figure}[ht]
\centerline{\epsfxsize=6cm\epsfbox{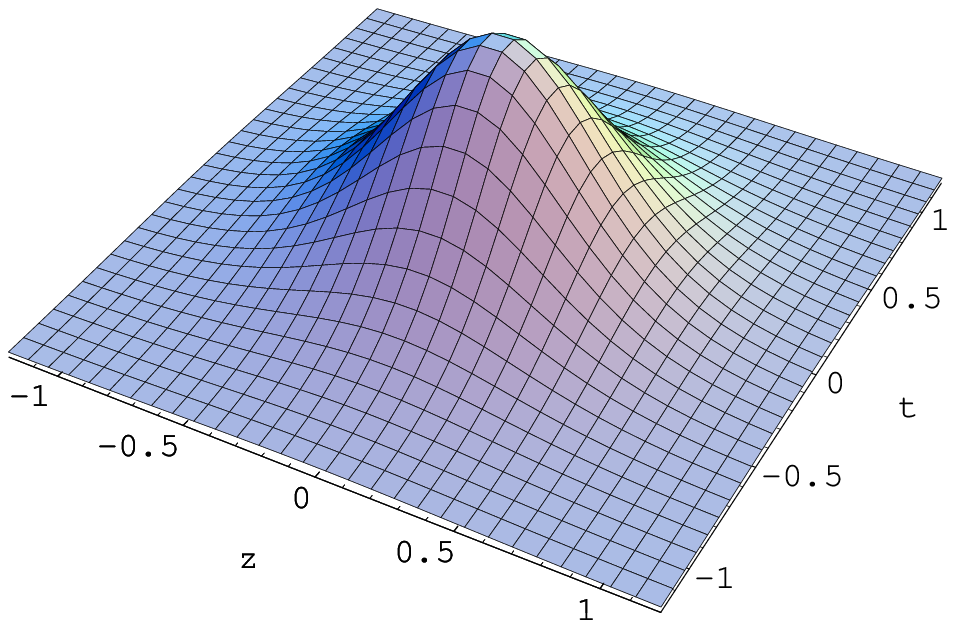}\epsfxsize=6cm\epsfbox{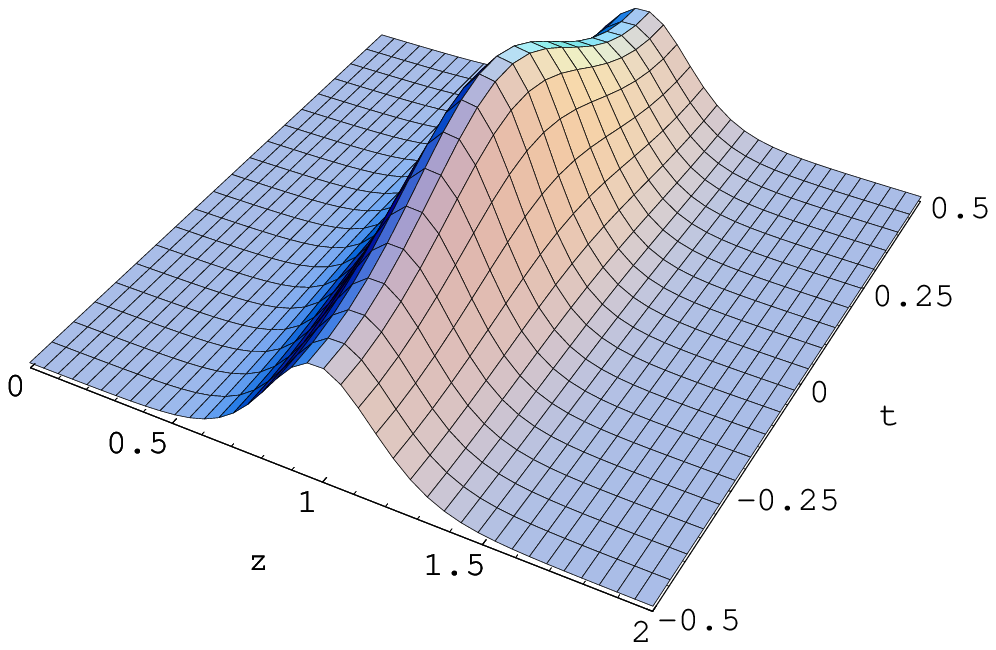}}   
\caption{Action densities of the ordinary BPST instanton (left) and of the periodic Harrington--Shepard
instanton with trivial holonomy (right) as function of $z,\;t$ at fixed $x=y=0$. The former is $O(4)$
and the latter is $O(3)$ symmetric. The size of the latter is $\rho=0.8\,\frac{1}{T}$.}
\end{figure}

In fact, the dependence on $\Lambda$ follows from the renormalization properties 
of the theory, therefore the caloron weight is a non-trivial function of one dimensionless 
variable, $\rho T$. At $\rho T\!\ll\!1$ it reduces to the 't Hooft's result for the 
ordinary BPST instanton~[18]. At $\rho T\!\gg\!1$ the caloron weight is~[9]

\beq
\int_0^{\frac{1}{T}} \!dz_4\int\!d^3z\!\int\!d^3U\!\int\!\frac{d\rho}{\rho^5}\,
\frac{C_0}{4\pi^2}\left(\frac{8\pi^2}{g^2(\mu)}\right)^4\,(\Lambda\rho)^{\frac{22}{3}}\,
(\pi\rho T)^{\frac{8}{3}}\,e^{-\frac{4}{3}(\pi\rho T)^2}. 
\la{GPY}\eeq
The last factor suppresses large calorons: it is the consequence of the Debye screening
mass which is nothing but the second derivative of the potential energy $P(A_4)$ at zero. 
 
The vacuum made of these calorons was built using the variational principle in Ref.~[23].
It turns out that the average of the Polyakov line $\half\!<\!\!L\!\!>$ grows rapidly 
from $0$ to $1$ near $T\approx \Lambda_{\overline{\rm MS}}$~[24], however 
strictly speaking, there is no mass gap and no confinement-deconfinement phase transition.

\section{Quantum weight of the caloron with non-trivial holonomy}
We define the quantum weight of the KvBLL caloron in the same way as it is done
in the case of ordinary instantons and the periodic instantons, see \eq{symbolic}.
The problem of writing down the Jacobian related to the caloron zero modes
has been actually solved already by Kraan and van Baal~[2]. 
Therefore, to find the quantum weight of the KvBLL caloron, only the ghost 
determinant needs to be computed. The KvBLL caloron has only the $O(2)$ symmetry 
corresponding to the rotation about the line connecting the dyon centers, 
and the determinant is a non-trivial function of three variables: 
the holonomy at spatial infinity encoded in the asymptotic value of 
$\left.\sqrt{A_4^aA_4^a}\right|_{|\vec x|\to\infty}\equiv \v$, the temperature T,
and the separation between the two dyons $\D$. Computing this function of three
variables looks like a formidable problem, however it has been solved
exactly in Ref.~[1].

We have followed Zarembo~[26] and first found the derivative of
the determinant $\Det(-D^2)$ with respect to the holonomy or, more precisely, to
$\v$. [The holonomy is $\half\Tr\,L=\cos(\v/2T)$]. The derivative 
$\partial\Det(-D^2)/\partial\v$ is expressed through the Green function of 
the ghost field in the caloron background~[26]. If a self-dual field is written 
in terms of the Atiyah--Drinfeld--Hitchin--Manin--Nahm
construction, and in the KvBLL case it basically is~[2,3], the Green function
is generally known~[27-29] and we build it explicitly for the KvBLL case. 
Therefore, we are able to find the derivative $\d\Det(-D^2)/\d\v$.
Next, we reconstruct the full determinant by integrating over $\v$
using the determinant for the trivial holonomy \ur{GPY} as a boundary condition.
This determinant at $\v=0$ is still a non-trivial function of the caloron size $\rho$ 
related to the dyon separation according to $r_{12}\!=\!|z_1\!-\!z_2|\!=\!\pi\,\rho^2T$,
and the fact that we match it from the $\v\neq 0$ side is a serious check.
Actually we need only one overall constant factor from Ref.~[9] in order
to restore the full determinant at $\v\neq 0$, and we make a minor improvement
of the Gross--Pisarski--Yaffe calculation as we have computed the needed
constant analytically. \\

Depending on the holonomy, the $M,L$ dyon cores are of the size $\frac{1}{\v}$ 
and $\frac{1}{\bv}$, respectively, where $\bv \!=\! 2\pi T-\v$. At large
dyon separations, $\D\!\gg\!1/T$, the 1-loop KvBLL caloron weight can be written
in a compact form in terms of the coordinates of the dyon centers~[1]:
\beqa
\nn
Z&= &\int d^3z_1 \, d^3z_2\, T^6\,C\left(\frac{8\pi^2}{g^2}\right)^4\!
\left(\frac{\Lambda e^{\gamma_E}}{4\pi T}\right)^{\frac{22}{3}}
\left(\frac{1}{T\D}\right)^{\frac{5}{3}}
\left(2\pi+\frac{\v \bv }{T}\D\right)\\
\nn\\
\nn
\!\!\!&\!\!\times \!\!&\!\left(\v\D\!+\!1\right)^{\frac{4\v}{3\pi T}\!-\!1}
\left(\bv\D\!+\!1\right)^{\frac{4\bv}{3\pi T}\!-\!1}
\exp\left[-V\,P(\v)-2\pi \D P''(\v)+...\right],\\
\nn\\
\nn
C&=&\frac{64}{\pi^2}\,\exp\left[{\frac{8}{9}-\frac{16\,{\gamma_E}}{3}
+\frac{2\pi^2}{27}+\frac{4\,{\zeta}'(2)}{\pi^2}}\right]= 1.031419972084\,,\\
\nn\\
\nn
P(\v)&=&\frac{1}{12\pi^2T}\v^2\bv^2,\qquad P''(\v)=\frac{d^2}{d\v^2}P(\v).
\eeqa
This expression is valid at $r_{12}T\!\gg\!1$ but arbitrary holonomy, $\v,\bv\!\in\![0,2\pi T]$, 
meaning that it is valid also for overlapping dyon cores.

\section{Dyon interaction}

At large separations, dyons (curiously) have the familiar ``linear plus Coulomb" 
interaction:
\beqa\nn
V(\D)&=&\D T^2 2\pi\left(\frac{2}{3}-4\:\nu(1\!-\!\nu)\right)\\
\nn
&-&\!\frac{1}{\D}\!\left(\frac{4}{3\pi}\log\left[\nu(1\!-\!\nu)(2\D T)^2\right]
+1.946\right)+\!\ldots,\quad
\nu\equiv \frac{\v}{2\pi T} \in [0,1].
\eeqa
This interaction is a purely {\em quantum effect}: classically dyons do not interact at all
since the KvBLL caloron is a classical solution whose action is independent of the dyon separation.

When the holonomy is not too far from trivial, $0.788<\half|\Tr\,L|<1$, such that $P''(\v)>0$,
dyons inside the KvBLL caloron attract each other, and calorons can be stable. At $\v\!\to\! 0$
this attraction is in fact the well-known effect of the suppression of large-size calorons 
owing to the non-zero Debye mass, cf. \eq{GPY}:
\beq\nn
\exp\left(-\frac{4\pi}{3}\,\D\,T\right) = \exp\left(-\frac{4}{3}\,(\pi\rho T)^2\right),
\eeq
More generally, the coefficient in the linear term is the second derivative of
the potential energy $P''(\v)$. Therefore, in the complementary range, $\half|\Tr\,L|<0.788$, 
where the second derivative changes sign, dyons experience a strong linearly rising 
repulsion, see Fig.~2. For these values of $\v$, integration over the dyon separation diverges: 
calorons with holonomy far from trivial ``ionize" into separate dyons.

\section{Caloron free energy and instability of the trivial holonomy}   

Let us make a crude estimate of the free energy of the non-interacting $N_+$ calorons 
and $N_-$ anti-calorons at small $\v<\pi T\left(1-\frac{1}{\sqrt{3}}\right)$ where the 
integral over dyon separation inside the KvBLL caloron converges:
\beqa\nn
Z_{\rm cal}&=&
\sum_{N_+,N_-}\frac{1}{N_+!N_-!}\left(\!\int\!d^3z\,T^3\,\zeta\right)^{N_++N_-}\;
\exp\left(-V\,P(\v)\right)\\
\nn\\
\nn
&&\!\!\!{\rm fugacity}\;\zeta\simeq \int_0^\infty\!d\D\,({\rm dyon\;weight})\,
\exp\left[-2\pi \D T\left(\frac{2}{3}-4\nu(1\!-\!\nu)\right)\right]\\
\nn\\
\nn 
&=& \exp\left[-VT^3 F(\nu,T)\right]
\eeqa
where
\beq\nn
F(\nu,T)=\frac{4\pi^2}{3}\nu^2(1\!-\!\nu)^2-2\,\zeta(T,\nu),\qquad \nu=\frac{\v}{2\pi T},
\eeq
is the free energy of the non-interacting caloron gas.

\begin{figure}[ht]
\centerline{\epsfxsize=8cm\epsfbox{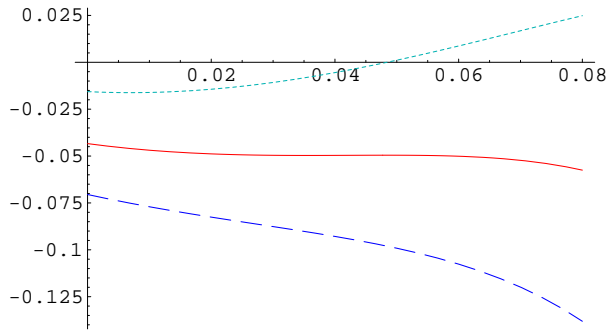}}   
\caption{Free energy of the caloron gas in units of $T^3V$ at
$T= 1.3\Lambda$ (dotted), $T= 1.125\Lambda$ (solid) and  $T= 1.05\Lambda$ (dashed)
as function of the asymptotic value of $A_4=\v$, in units of $2\pi T$.}
\end{figure}

At high temperatures the free energy is dominated by the perturbative energy $P(A_4)$.
Calorons are lowering the free energy but their non-perturbative effect is small.
Therefore, the minimum corresponds to the holonomy being close to the trivial one. 

However, below the critical temperature 
$T_c=1.125\,\Lambda$ trivial holonomy becomes unstable, and the system rolls to large 
$\v$ where dyons repulse each other. To find the free energy at large values of $\v$,
one has to consider the statistical mechanics of the interacting system of 
$L,M,\overline L,\overline M$ dyons carrying all four possible combinations of the electric
and magnetic charges. This has not been done. One can expect, however, that
the minimum of the full free energy below $T_c$ will occur at $\v=\pi T$
corresponding to $\Tr L=0$. Probably it will mean confinement, with the
correlator of two Polyakov lines decaying as an exponent of the separation times
the second derivative of the free energy at $\v=\pi T$, and with the area
law for the spatial Wilson loop determined by the magnetic screening length.

\section{Summary}
\noi 1. The KvBLL caloron is ``made of" two Bogomolnyi--Prasad--Sommerfeld monopoles 
or dyons and characterized by non-trivial holonomy, $\half\,\Tr\,L\neq 1,-1$ (in $SU(2)$). 
It is self-dual, and has one unit of topological charge. \\
\vskip -0.2true cm

\noi 2. The quantum weight, or the probability with which KvBLL calorons appear in the
YM partition function has been computed {\em exactly} at 1-loop.\\
\vskip -0.2true cm

\noi 3. At large separation of constituent dyons, they experience a linear rising
{\em attraction} if $\half\,|\Tr\,L|\approx 1$ or {\em repulsion}
if $\half\,|\Tr\,L|\approx 0$. \\
\vskip -0.2true cm

\noi 4. At very high temperatures only calorons with trivial holonomy survive
($\half\,|\Tr\,L|=1$). At temperatures below critical $T_c\simeq 1.125\,\Lambda$
trivial holonomy becomes unstable, and calorons ``ionize" into separate dyons. 
It may be that at this point the confinement-deconfinement transition takes place.

\end{document}